\documentclass[12pt]{iopart}
\newcommand{\beq}[1]{\begin{equation}\label{#1}}
\newcommand{\eeq}{\end{equation}}
\newcommand{\bear}[1]{\begin{eqnarray}\label{#1}}
\newcommand{\ear}{\end{eqnarray}}
\newcommand{\nn}{\nonumber}
\newcommand{\R}{\mbox{\rm I$\!$R}}

\newcommand{\eps}{\varepsilon}
\begin{document}
\title[Friedmann universe with perfect fluid and scalar field]
{General solutions for  flat Friedmann universe
filled by perfect fluid and scalar field with exponential potential}

\author{Hienz Dehnen\dag, V R Gavrilov\ddag\footnote[3]{To whom
correspondence should be addressed (gavr@rgs.phys.msu.su)} and
V N Melnikov\ddag}

\address{\dag\ Universit$\ddot a$t Konstanz, Fachbereich Physik, Fach M 568
         D-78457 Konstanz, Germany}
\address{\ddag\ Centre for
Gravitation and Fundamental Metrology, VNIIMS, and Institute for Gravitation
and Cosmology, PFUR, 3-1 M.Ulyanovoy St., Moscow 117313, Russia}
\eads{\mailto{Heinz.Dehnen@uni-konstanz.de},
\mailto{gavr@rgs.phys.msu.su} and
\mailto{melnikov@rgs.phys.msu.su}}

\begin{abstract}
We study integrability by quadrature of a spatially flat Friedmann model
containing both a minimally coupled scalar field $\varphi$ with an
exponential potential
$V(\varphi)\sim\exp[-\sqrt{6}\sigma\kappa\varphi]$,
$\kappa=\sqrt{8\pi G_N}$, of arbitrary sign
and a perfect fluid with barotropic equation of
state $p=(1-h)\rho$. From the mathematical view point the model is
pseudo-Euclidean Toda-like system with 2 degrees of freedom. We
apply the methods developed in our previous papers, based on the
Minkowsky-like geometry for 2 characteristic vectors
depending on the parameters $\sigma$ and $h$.
In general case the problem
is reduced to integrability of a
second order ordinary differential equation known as the
generalized Emden-Fowler equation, which was  investigated by
discrete-group methods. We present 4 classes of general solutions
for the parameters obeying the following relations: {\bf A}.
$\sigma$ is arbitrary, $h=0$; {\bf B}.  $\sigma=1-h/2$, $0<h<2$;
{\bf C1}. $\sigma=1-h/4$, $0<h\leq 2$; {\bf C2}. $\sigma=|1-h|$,
$0<h\leq 2$, $h\neq 1,4/3$. We discuss the properties of the exact
solutions near the initial singularity and at the final stage of
evolution.
\end{abstract}

\submitto{\CQG}
\pacs{04.20.Jb}

\maketitle
\section{Introduction}
A large class of theories (multidimensional
\cite{Mel1,Mel2,Mel3}, Kaluza-Klein models, supergravity and
superstring theories, higher order gravity, see for instance
\cite{FJ98},\cite{T01}) deal with a weakly coupled scalar field
$\varphi$ with a potential of the form
\beq{1.1}
 V(\varphi)=\frac{V_0}{\kappa^2}\rme^{-\sqrt{6}\sigma\kappa\varphi},
\eeq
where $\kappa=\sqrt{8\pi G_N}$, $\sigma$ is a dimensionless positive
constant, characterizing the steepness of the potential, and the constant
$V_0$ may be positive and negative.  A number of authors
\cite{FJ98}-\cite{R01} (see also references therein) have studied a
spationally homogeneous and isotropic Friedmann model containing both a
scalar field and a perfect fluid subject to the linear equation of state
\beq{1.2}
p=(1-h)\rho,
\eeq
where the constant $h$ satisfied $0\leq h \leq 2$. The attention was mainly
focussed on the qualitative behaviour of solutions, stability of the
exceptional solutions to curvature and shear perturbations and their
possible applications within the known cosmological scenaria such as
inflation and scaling ("tracking") . In particular, it was found by a
phase plane analysis \cite{W98,B98,W99} that for "flat"
positive potentials ($V_0>0,\ 0<\sigma^2<1-h/2$) there exists an unique
late-time attractor in the form of the scalar dominated solution. It is
stable within homogeneous and isotropic models with non-zero spatial
curvature with respect to spatial curvature
perturbations for $\sigma^2<1/3$ and provides the power-law inflation. For
"intermediate" positive potentials ($V_0>0,\ 1-h/2<\sigma^2<1$) an unique
late-time attractor is the  scaling solution, where the
scalar field "mimics" the perfect fluid, adopting its equation of state.
The energy-density of the scalar field scales with that of the perfect
fluid.
For $h>4/3$
this solution is  stable within generic Bianchi models
to curvature and shear perturbations
and provides  the power-low inflation.
The scaling solution is unstable to curvature perturbations, when
$0<h<4/3$, although it is stable to shear perturbations.
Regions on $(\sigma^2,h)$
parametrical plane corresponding to various qualitative evolution
are presented on figure 1.
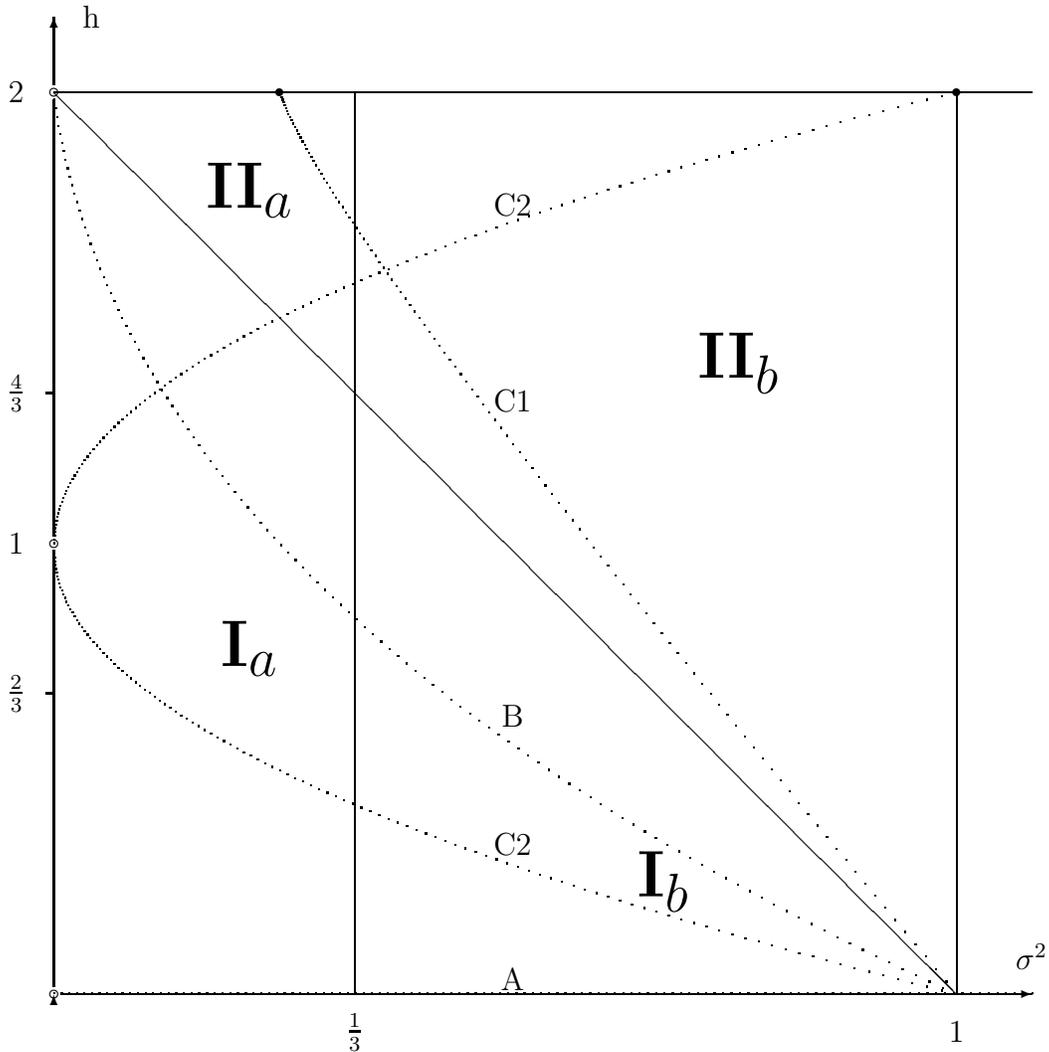
\begin{figure}
\begin{center}
\unitlength=1.00mm
\special{em:linewidth 0.4pt}
\linethickness{0.4pt}
\begin{picture}(140.00,140.00)
\put(10.00,140.00){\vector(0,0){0.00}}
\put(10.00,10.00){\vector(0,0){0.00}}
\put(15.00,140.00){\makebox(0,0)[cc]{h}}
\put(140.00,15.00){\makebox(0,0)[cc]{$\sigma^2$}}
\put(10.00,130.00){\line(1,0){130.00}}
\put(130.00,130.00){\line(0,-1){120.00}}
\put(10.00,130.00){\line(1,-1){120.00}}
\put(50.00,10.00){\line(0,1){120.00}}
\put(10.00,130.00){\circle{1.00}}
\put(11.00,10.00){\vector(1,0){129.00}}
\put(10.00,70.00){\circle{1.00}}
\put(10.00,11.00){\line(0,1){58.00}}
\put(10.00,71.00){\line(0,1){58.00}}
\put(10.00,131.00){\vector(0,1){9.00}}
\put(5.00,70.00){\makebox(0,0)[cc]{1}}
\put(5.00,90.00){\makebox(0,0)[cc]{$\frac{4}{3}$}}
\put(130.00,5.00){\makebox(0,0)[cc]{1}}
\put(50.00,5.00){\makebox(0,0)[cc]{$\frac{1}{3}$}}
\put(40.00,130.00){\circle*{1.00}}
\put(130.00,130.00){\circle*{1.00}}
\put(10.00,10.00){\circle{1.00}}
\put(5.00,130.00){\makebox(0,0)[cc]{2}}
\put(5.00,50.00){\makebox(0,0)[cc]{$\frac{2}{3}$}}
\put(9.00,90.00){\line(1,0){1.00}}
\put(9.00,50.00){\line(1,0){1.00}}
\bezier{120}(10.00,70.00)(10.00,100.00)(130.00,130.00)
\bezier{120}(10.00,69.00)(10.00,40.00)(130.00,10.00)
\bezier{120}(40.00,130.00)(46.00,111.00)(130.00,10.00)
\bezier{120}(10.00,129.00)(20.00,60.00)(130.00,10.00)
\bezier{120}(10.00,10.20)(70.00,10.20)(140.00,10.20)
\put(71.00,115.00){\makebox(0,0)[cc]{C2}}
\put(71.00,89.00){\makebox(0,0)[cc]{C1}}
\put(71.00,47.00){\makebox(0,0)[cc]{B}}
\put(71.00,30.00){\makebox(0,0)[cc]{C2}}
\put(71.00,12.00){\makebox(0,0)[cc]{A}}
\put(36.00,117.00){\makebox(0,0)[cc]{\bf\Huge II$_a$}}
\put(36.00,56.00){\makebox(0,0)[cc]{{\bf\huge I$_a$}}}
\put(101.00,94.00){\makebox(0,0)[cc]{{\bf\Huge II$_b$}}}
\put(91.00,25.00){\makebox(0,0)[cc]{{\bf\Huge I$_b$}}}
\end{picture}
\end{center}
\caption{
The domains {\bf Ia, Ib, IIa and IIb} are bounded by the
lines $h=0,\ h=2,\ \sigma^2=0,\ \sigma^2=1/3,\ \sigma^2=1-h/2$. The
solutions corresponding to {\bf Ia and IIa} are inflationary on
late times with an attractor stable to curvature perturbations.
The late-time attractor for {\bf Ia and Ib} is the scalar
dominated solution and for {\bf IIa and IIb} it is the scaling
solution. The dotted curves {\bf A,B,C1} and {\bf C2 } are identified in
the text and present the integrable by quadrature classes of the
model.}
\end{figure}

Integrability of the model is not well studied yet.
Only  for the special case for $h=1$ (dust) and $\sigma =1/2$
the procedure of getting a solution to the model  was
given in \cite{R01}.

In the present paper we study the problem of integrability by quadrature
of a spatially flat Friedmann model containing both a minimally coupled
scalar field with the  exponential potential (\ref{1.1}) and a perfect
fluid with the equation of state  (\ref{1.2}) for several classes of sets
of $(\sigma, h)$-parameters.  We apply the methods developed in our
previous papers \cite{IM94}-\cite{GM98} devoted to integrability of
multidimensional cosmological models.  It is clear that the possibility of
reducing the problem to quadrature  depends on the parameters $\sigma$
and $h$. We show that in a general case the problem is reduced to
integrability of a second order ordinary differential equation known as
the generalized Emden-Fowler equation, which was  investigated by
discrete-group methods \cite{PZ},\cite{ZP}.  We present here 4 classes of
general solutions for the parameters obeying the following relations.

{\bf A}. $\sigma$ is arbitrary, $h=0$.

{\bf B}.  $\sigma=1-h/2$, $0<h<2$.

{\bf C1}. $\sigma=1-h/4$, $0<h\leq 2$.

{\bf C2}. $\sigma=|1-h|$, $0<h\leq 2$, $h\neq 1,4/3$.\\
The corresponding curves on $(h,\sigma^2)$ parametrical plane are presented
on figure 1.

The paper is organized as follows. In section 2 we describe the
model and obtain the Einstein-scalar field equations in
the form of the Lagrange-Euler equations following from some Lagrangian.
Dynamical system described by the Lagrangian of this form belongs to the
class of pseudo-Euclidean Toda-like systems. To integrate them we apply in
section 3 the methods developed in our papers \cite{IM94,GIM95,GM98} on
multidimensional cosmology.  The method used  in the general case {\bf C}
($\sigma^2\neq (1-h/2)^2,\  0<h\leq 2$) is based on reducing the
Euler-Lagrange equations to the generalized  Emden-Fowler (second-order
ordinary differential) equation. To conclude the paper, we discuss in
section 4 the physical properties of the obtained exact solutions.
Asymptotic behaviour of these solutions at early and late times  is
analyzed. Example of the explicit solution in the cosmic time is
presented.

\section{The model}
We start with Einstein Eqs. in a spatially flat Friedmann metric
\beq{2.1}
\rmd s^2=-\rme ^{2\alpha(t)}\rmd t^2  +
\rme ^{2x(t)}
 \left[\rmd r^2 + r^2\left(\rmd \theta^2+\sin^2\theta{\rm
 d}\phi^2\right)\right], \end{equation}
where $\exp[x(t)]\equiv a(t)$ is the scale factor and the function
$\alpha(t)$ determines a time gauge ($\alpha(t)\equiv 0$
corresponds to the cosmic time $t_c$ gauge). We assume that the
universe contains  both a self-interacting scalar field $\varphi$
with a potential $V(\varphi)$ and a separately conserved perfect
fluid with the barotropic equation of state (\ref{1.2}). The
governing set of equations, which follows from coupled Einstein
and scalar field equations, reads
\bear{2.2} \dot
x^2=\frac{\kappa^2}{3}
\left[ \frac{1}{2}\dot\varphi^2 +
\left( V(\varphi)+\rho\right)\rme^{2\alpha}\right],\\
\label{2.3}
\ddot x= \frac{1}{2}(\dot\alpha-3\dot x)\dot x - \frac{\kappa^2}{2}
\left[ \frac{1}{2}\dot\varphi^2 -
\left( V(\varphi) - p\right)\rme^{2\alpha}\right],\\
\label{2.4}
\ddot\varphi=(\dot\alpha-3\dot x)\dot\varphi - V'(\varphi)\rme^{2\alpha},\\
\label{2.5}
\dot\rho=-3\dot x(p + \rho).
\ear
Using the barotropic equation of state we get immediately
>from Eq.(\ref{2.5})
\beq{2.6}
\rho=\frac{\rho_0}{\kappa^2}\rme^{-3(2-h)x},
\eeq
where $\rho_0$ is an arbitrary positive constant.

It can be easily verified, that the set of Eqs.(\ref{2.2})-(\ref{2.4}),
where the presence of a perfect fluid density $\rho$ and its pressure $p$
is cancelled by Eqs.(\ref{1.2})and (\ref{2.6}),
is equivalent to the set of Euler-Lagrange equations obtained from the
Lagrangian
\beq{2.7}
L(x,y,\alpha,\dot x,\dot y)=\frac{1}{2}\rme^{3x-\alpha}
\left( - \dot x^2 + \dot y^2 \right)
- \frac{\kappa^2}{6}\rme^{\alpha-3x}
\left[ \rho_0\rme^{3hx} +
\rme^{6x}V\left(\frac{\sqrt{6}}{\kappa} y\right)\right],
\eeq
where we introduced the following dimensionless variable
\bear{}
y=\frac{\kappa}{\sqrt{6}}\varphi.\nn
\ear
The equation
$\partial L/\partial\alpha=\rmd (\partial L/\partial\dot\alpha)/\rmd t=0$
leads to the constraint Eq.(\ref{2.2}).
 Fixing the gauge $\alpha\equiv F(x,y)$, we can consider
 Eqs.(\ref{2.3}),(\ref{2.4}) as the Euler-Lagrange equations obtained from
the Lagrangian (\ref{2.7}) under the zero-energy constraint (\ref{2.2}).

In what follows we consider the potential of the form (\ref{1.1}).
As the system is symmetric under the transformation $\sigma\to
-\sigma,\ \varphi\to -\varphi$, without a loss of generality we
will consider only the case $\sigma >0$. For the exponential
potential (\ref{1.1}) the Lagrangian (\ref{2.7}) looks as follows
\beq{2.8}
L(x,y,\alpha,\dot x,\dot y)=\frac{1}{2}\rme^{3x-\alpha}
\left( - \dot x^2 + \dot y^2 \right) -
\frac{\kappa^2}{6}\rme^{\alpha-3x} \left[ \rho_0\rme^{3hx} +
V_0\rme^{6(x-\sigma y)}\right]. \eeq Dynamical system described by
the Lagrangian of this form belongs to the class of
pseudo-Euclidean Toda-like systems investigated in our previous
papers \cite{IM94,GIM95,GM98}. Methods for integrating of
pseudo-Euclidean Toda-like systems are based on the Minkowski-like
geometry for characteristic vectors
$(\alpha_0+3(h-1),0)\in \R^2$ and
$(\alpha_0+3,-6\sigma)\in \R^2$
appearing when one puts the gauge $\alpha=\alpha_0 x$ with
$\alpha_0=$const.
Here we do not
describe the methods and refer to the above mentioned papers.
\section{General solutions}
{\bf A}. $\sigma$ is arbitrary, $h=0$.\\
Here we suppose that the perfect fluid pressure $p$ is equal to
its density $\rho$
(Zeldovich-type, or stiff matter). In this case the system
with Lagrangian (\ref{2.8}) is
integrable for arbitrary parameter $\sigma$ in the so-called harmonic
time gauge
defined by
\beq{3.1}
\alpha(x)=3x.
\eeq
Let us consider two different cases: $\sigma\neq 1$ and $\sigma=1$.

If $\sigma\neq 1$ we introduce the following variables
\bear{}
u=-\sigma x + y, \ \ v=x - \sigma y. \nn
\ear
In the terms of $u$ and $v$ the equations of motion and the zero-energy
constraint look as follows
\bear{}
\ddot u =0,\nn\\
\ddot v = V_0(1-\sigma^2)\rme^{6v}, \nn\\
-\dot u^2 + \dot v^2 = \frac{1-\sigma^2}{3}
\left[ \rho_0 + V_0\rme^{6v} \right].\nn
\ear
We notice that for $v$ we get the Liouville equation. Integrating this set
of equations and
inverting the linear transformation, we get the following general solution:

the scale factor
\beq{3.2}
a\equiv\rme^x= a_0
\left[ f(t-t_0)\rme^{\sigma A(t-t_0)}\right]^{1/[3(\sigma^2-1)]},
\eeq

the scalar field
\beq{3.3}
\varphi=\frac{\sqrt{6}}{\kappa}
\left\{
\ln\left[ f^{\sigma}(t-t_0)\rme^{A(t-t_0)}\right]^{1/[3(\sigma^2-1)]}+ y_0
\right\},
\eeq
where we introduced the function
\bear{}
f(t-t_0)&&=\sinh(\sqrt{B}|t-t_0|)/\sqrt{B}, \  V_0(1-\sigma^2)>0,\ B>0,\nn\\
&&=\cosh(\sqrt{B}|t-t_0|)/\sqrt{B}, \  V_0(1-\sigma^2)<0,\ B>0,\nn\\
&&=\sin(\sqrt{B}|t-t_0|)/\sqrt{B}, \  V_0(1-\sigma^2)>0,\ B<0,\nn\\
&&=|t-t_0|, \  V_0(1-\sigma^2)>0,\ B=0.\nn
\ear
The constant $B$ is defined by
\bear{}
B=A^2 + 3(1-\sigma^2)\rho_0.\nn
\ear
The general solution contains 2 arbitrary constants $t_0$, $A$ and 2
constants $a_0$,$y_0$, obeying the constraint:
\bear{}
\rme^{6\sigma y_0}=3a^6_0|V_0(1-\sigma^2)|.\nn
\ear

For the second case
$\sigma=1$ the equations of motion and the zero-energy constraint with
respect to the harmonic time gauge read in the old variables
\bear{}
\ddot x= V_0\rme^{6(x-y)},\nn\\
\ddot y= V_0\rme^{6(x-y)},\nn\\
-\dot x^2 + \dot y^2 = -\frac{1}{3}
\left[ \rho_0 + V_0\rme^{6(x-y)} \right].\nn
\ear
We immediately find the integral of motion
$\dot x - \dot y={\rm const}\equiv2A$.

If $V_0>0$ the constant $A$ is nonzero due to the
zero-energy constraint.
The general solution in this case for arbitrary $V_0$ looks as follows
\bear{3.4}
a\equiv \rme ^x=\exp
\left\{
\left(   A+ \frac{\rho_0}{12A}   \right)(t-t_0) +
V_0\left[\rme^{12A(t-t_0)}-1\right] +
x_0
\right\},\\
\label{3.5}
\varphi=\frac{\sqrt{6}}{\kappa}
\left\{
\left(-A+ \frac{\rho_0}{12A}\right)(t-t_0) +
V_0\left[\rme^{12A(t-t_0)}-1\right] +
y_0
\right\},
\ear
where the constants $A,t_0$ are arbitrary and the constants $x_0,y_0$ obey
the relation
\linebreak
$y_0-x_0=2At_0$.

For $V_0<0$ the additional solution appears (corresponding to $A=0$,
$\dot x=\dot y$)
\bear{}
a\equiv\rme^x=\exp
\left\{
B(t-t_0) -\frac{\rho_0}{2}(t-t_0)^2+x_0
\right\},\nn\\
\varphi=\frac{\kappa}{\sqrt{6}}
\left[
\ln a + \frac{1}{6}\ln(-\rho_0/V_0)
\right],\nn
\ear
where $B$ and $x_0$ are arbitrary constants.
We remind that $t$ is the harmonic time. It is connected
with the cosmic time $t_c$ by the differential equation $\rmd t_c=a^3\rmd
t$.

{\bf B}.  $\sigma=1-h/2$, $0<h<2$.\\
We notice that the model for $\sigma=1/2$ and $h=1$ has been integrated in
\cite{R01}. Here we study a more general case. We fix the time gauge
choosing
the following function $\alpha(x)$:
\bear{}
\alpha(x)=3(1-h)x.\nn
\ear
Then, with respect to the new variables $u$ and $v$ defined by
\bear{}
u=\exp\left[ \frac{3h}{2}(x-y)\right],\ \
v=\exp\left[ \frac{3h}{2}(x+y)\right]\nn
\ear
the equations of motion and the zero-energy constraint may be written in a
rather simple form
\bear{}
\ddot u =0,\nn\\
\ddot v = \frac{3}{2} h(2-h) V_0 u^{(4-3h)/h},\nn\\
\dot u\dot v = \frac{3}{4}h^2\left[ \rho_0 + V_0 u^{2(2-h)/h}\right].\nn
\ear

The set of equations is easily integrable. We obtain
\bear{}
u=A(t-t_0)>0,\nn\\
v=\frac{3h^2}{4A^2}
\left\{
\rho_0A(t-t_0) + \frac{h}{4-h}V_0[A(t-t_0)]^{(4-h)/h} + B
\right\}>0,\nn
\ear
 where $A$ is an arbitrary nonzero constant  and $B$ is an arbitrary
 nonnegative constant, $V_0$ has arbitrary sign.
 If $V_0<0$, then, the following additional special
 solution arises

\bear{}
u=(-\rho_0/V_0)^{h/[2(2-h)]},\nn\\
v=\frac{3}{4} h(2-h) \rho_0^{h/(2-h)]}
\left[ T^2 - (t-t_0)^2 \right],\nn
\ear
where the integration constant $T\neq 0$.
Then, one easily gets the scale factor
\bear{3.6}
a\equiv \rme^ x= (uv)^{1/(3h)}
\ear
and the scalar field
\bear{3.7}
\varphi=\frac{\sqrt{2/3}}{\kappa h}\ln \frac{v}{u}.
\ear

{\bf C}. $\sigma^2\neq (1-h/2)^2,\  0<h\leq 2$.\\
We introduce the following variables for the factorization of the
potential in the
Lagrangian (\cite{GM98})
\bear{}
u=3[\sigma x - (1-h/2)y],\nn\\
v=3[(h/2-1)x + \sigma y] - \ln\sqrt{|V_0|/\rho_0}.\nn
\ear
Then, the equations of motion and the zero-energy constraint for
$u$ and $v$ in
the harmonic time gauge defined by Eq.(\ref{3.1}) look as follows
\bear{3.8}
\ddot u=\frac{3}{2}h\sigma A_0
\left(\rme^{2v} + \eps\right)
\exp
\left\{
\frac
{h\sigma u  +  (2-h-2\sigma^2)v}
{\sigma^2\ - (1-h/2)^2}
\right\},\\
\label{3.9}
\fl
\ddot v=-\frac{3}{2} A_0
\left(h(1-\frac{h}{2})\rme^{2v} + (2-h-2\sigma^2)\eps\right)
\exp
\left\{
\frac
{h\sigma u  +  (2-h-2\sigma^2)v}
{\sigma^2\ - (1-h/2)^2}
\right\},\\
\label{3.10}
\fl
-\dot u^2 + \dot v^2 +
3A_0[\sigma^2\ - (1-h/2)^2]\left(\rme^{2v} + \eps\right)
\exp
\left\{
\frac
{h\sigma u  +  (2-h-2\sigma^2)v}
{\sigma^2\ - (1-h/2)^2}
\right\}
=0,
\ear
where we denoted
\bear{}
A_0=
\rho_0^
{-\frac{2-h-2\sigma^2}{2[\sigma^2\ - (1-h/2)^2]}}
|V_0|^
{\frac{h(1-h/2)}{2[\sigma^2\ - (1-h/2)^2]}},\nn\\
\eps={\rm sgn}(V_0).\nn
\ear
Let us express $\dot v^2$ from the zero-energy condition
(\ref{3.10}) as follows
\bear{3.11}
\fl
\dot v^2=
\left[\left( \frac {\rmd u}{\rmd v}\right)^2-1\right]^{-1}
3A_0[\sigma^2\ - (1-h/2)^2]\left(\rme^{2v} + \eps\right)
\exp
\left\{
\frac
{h\sigma u  +  (2-h-2\sigma^2)v}
{\sigma^2\ - (1-h/2)^2}
\right\}.
\ear
Substituting $\ddot u$, $\ddot v$ and $\dot v^2$
into the relation
\bear{3.12}
\frac {\rmd ^2u}{\rmd ^2v}=
\frac {\ddot u - \ddot v \rmd u/\rmd v}{\dot v^2},
\ear
we obtain the following second order ordinary differential equation
\bear{3.13}
\fl
\frac{\rmd ^2u}{\rmd v^2}=
\left[\left( \frac{\rmd u}{\rmd v}\right)^2-1\right]
\left\{
\frac{1}{2}\left(-\frac{\sigma^2\ - (1-h^2/4)}{\sigma^2\ - (1-h/2)^2}+
\frac{\rme ^{2v}-\eps }{\rme ^{2v}+\eps }\right)
\frac{\rmd u}{\rmd v}+ \frac{h\sigma/2}{\sigma^2\ - (1-h/2)^2}
\right\}.
\ear
The procedure is valid if $\dot v\not\equiv 0$.

The exceptional solution with $\dot v\equiv 0$ appears only for the positive
potential when
$\sigma^2>1-h/2$ and $0<h<2$. In the terms of the cosmic time $t_c$ it reads
\bear{3.14}
a= \left\{
\sqrt{\frac{3\rho_0}{4(\sigma^2-(1-h/2)}} \sigma(2-h)|t_c-t^0_c|
\right\}^{2/[3(2-h)]},\\
\label{3.15}
\varphi=\frac{\sqrt{2/3}}{\kappa\sigma}\ln
\left(
\sqrt{\frac{3}{h}(2-h)V_0} \sigma |t_c-t^0_c|
\right).
\ear

It should be mentioned, that the set of the equations
(\ref{3.8})-(\ref{3.10})
does not admit static solutions $\dot u=\dot v\equiv 0$
as well as the solutions with $\dot u=\pm\dot v$.
So, using the relations
(\ref{3.11}) and (\ref{3.12})
we do not lose any solutions
except, possibly, the exceptional solution (\ref{3.14}),(\ref{3.15}).

Let us suppose that one is able to obtain the general solution  to the
equation (\ref{3.13})
 in the parametrical form $v=v(\tau)$, $u=u(\tau)$,
where $\tau$ is a parameter.
Then, we obtain the scale factor $a\equiv\exp[x]$ and the scalar field
$\varphi= (\sqrt{6}/\kappa)y$
as  functions of the parameter $\tau$.
The relation between the parameter $\tau$ and the harmonic time $t$ may be
always derived by integration of the zero-energy constraint written in the
form of the following separable equation
\bear{3.16}
\fl
\rmd t^2=\frac
{\left(\frac{\rmd u}{\rmd \tau}\right)^2-
\left(\frac{\rmd v}{\rmd \tau}\right)^2}
{3A_0[\sigma^2\ - (1-h/2)^2]\left(\rme^{2v} + \eps\right)}
\exp
\left\{
-\frac
{h\sigma u  +  (2-h-2\sigma^2)v}
{\sigma^2\ - (1-h/2)^2}
\right\}
\rmd \tau^2.
\ear

Thus, the problem of the integrability by quadrature
of the model
is reduced to the integrability of the equation (\ref{3.13}).
For $\rmd u/\rmd v$  it represents the first-order
nonlinear ordinary differential equation. Its right hand side is the
third-order
polynomial (with coefficients depending on $v$) with respect to the
$\rmd u/\rmd v$. An equation of such type is called Abel's equation
(see, for instance, \cite{PZ},\cite{ZP}).

First of all let us
notice that the equation (\ref{3.13}) has the partial solution
$u=\pm v+$const
that make the relation (\ref{3.11}) singular. However, as was
already mentioned,
the set of Eqs.(\ref{3.8})-(\ref{3.10}) does not admit
the solutions with $\dot u=\pm\dot v$.
Existence of
this partial solution  to the Abel equation (\ref{3.13}) allows one
to find the
following nontrivial transformation
\bear{3.17}
\rme ^{2v}&=&-\eps \frac{X}{Y}\frac{\rmd Y}{\rmd X},\\
\label{3.18}
u&=&\delta\left[
v+\ln\left|\frac{Y}{X}\right| +\ln C\right], \ \ \delta=\pm 1,\ \ C>0,
\ear
that reduces it to the so-called generalized
Emden-Fowler equation
\bear{3.19}
\frac{\rmd ^2Y}{\rmd X^2}={\rm sgn}[\sigma^2\ - (1-h/2)^2]
\left(-\eps \frac{\rmd Y}{\rmd X}\right)^l Y^m X^n,
\ear
where the constant parameters $l,m$ and $n$ read
\bear{3.20}
l=\frac
{2(\delta\sigma - 1 + h/4)}
{\delta\sigma - 1 + h/2},\
m=-\frac
{\delta\sigma + 1 - h}
{\delta\sigma + 1 - h/2},\
n=-m-3.
\ear
In the special case $l=0$ Eq.(\ref{3.19})  is known as the Emden-Fowler
equation.

There are no methods for
integrating of the generalized Emden-Fowler equation with arbitrary
independent parameters $l,m$ and $n$.
However, the discrete-group  methods
developed in \cite{ZP} allow to integrate by
quadrature 3 two-parametrical classes, 11 one-parametrical classes and about
90 separated points in the parametrical space $(l,m,n)$ of the generalized
Emden-Fowler equation. Further we consider the following integrable classes.

{\bf C1}. $\sigma=1-h/4$, $0<h\leq 2,\ \delta=1$.\\
The parameters $l$ and $m$ given by Eq. (\ref{3.20}) are the
following
\bear{LM1}
l=0,\ m=-1 + \frac{2h}{8-3h}\in (-1,1]. \ear
The general solution to the generalized Emden-Fowler equation
(\ref{3.19}) with these parameters reads
\bear{}
Y=\frac{\tau}{F_m(\tau)}>0,\ \ X=\frac{1}{F_m(\tau)}>0,\nn \ear
where we introduced the following function \bear{3.21}
F_m(\tau)=\pm\int\left[\frac{2}{m+1}\tau^{m+1}+C_1\right]^{-1/2}
\rmd \tau + C_2. \ear The variable $\tau$ changes on the
interval which follows from $G_m(\tau)>0$, where we used the
function \bear{3.22} G_m(\tau)= \eps \left[ \frac{F_m(\tau)}{\tau
F_m^{\prime}(\tau)}-1 \right] \ear
 equal to the right hand side of Eqs.(\ref{3.17}) with substitutions of
 $X,Y$ and (\ref{3.21}).
Finally, using Eqs.(\ref{3.17}),(\ref{3.18}) we
find the scale factor
\bear{3.23}
a=a_0 \tau^{\frac{4(4-h)}{3h(8-3h)}}
G_m^{\frac{2}{3h}}(\tau)
\ear
and the scalar field
\bear{3.24}
\varphi=\frac{\sqrt{6}}{\kappa}
\left\{
 \ln \left[\tau^{\frac{8(2-h)}{3h(8-3h)}}
 G_m^{\frac{2}{3h}}(\tau)\right]+ y_0 \right\}, \ear
where
\bear{}
a_0= C^{\frac{4(4-h)}{3h(8-3h)}} (|V_0|/\rho_0)^{\frac{4(2-h)}{3h(8-3h)}},\
y_0=\ln
\left[
C^{\frac{8(2-h)}{3h(8-3h)}} (|V_0|/\rho_0)^{\frac{2(4-h)}{3h(8-3h)}}
\right].\nn
\ear
The relation between the variable $\tau$ and the cosmic time $t_c$
is the following
\bear{3.25}
\rmd  t_c^2= I_0
\tau^{\frac{4(2-h)(4-3h)}{h(8-3h)}}
G^{\frac{2(2-h)}{h}}_m(\tau)
\left[ F_m^{\prime}(\tau)\right]^2 \rmd \tau^2,
\ear
where
\bear{}
I_0= 16
\rho_0^{-\frac{(4-h)^2}{h(8-3h)}}
|V_0|^{\frac{4(2-h)^2}{h(8-3h)}}
C^{\frac{4(2-h)(4-h)}{h(8-3h)}}/[3h(8-3h].\nn
\ear
The general solution (\ref{3.23}),(\ref{3.24})
(with $l$ and $m$ given by(\ref{LM1}))
contains 3 arbitrary
constants $C$, $C_1$ and $C_2$  as required.

{\bf C2}. $\sigma=|1-h|$, $0<h\leq 2$,
$h\neq 1,4/3, \delta={\rm sgn}(1-h)$.\\
Now we have the following parameters in the generalized Emden-Fowler
equation
\bear{LM2}
l=3,\ \ m=-1 + \frac{h}{4-3h}.
\ear
Its general solution looks as follows
\bear{}
X=\frac{\tau}{R_m(\tau)}>0,\ \ Y=\frac{1}{R_m(\tau)}>0,\nn
\ear
where the function $R_m(\tau)$ is defined by
\bear{3.26}
R_m(\tau)&=&\pm\int\left[\frac{-2\eps}{|m+2|}\tau^{-m-2}+C_1\right]^{-1/2}
\rmd \tau + C_2, \ \ m\neq -2,\\
\label{3.27}
&=&\pm\int\left[\eps\ln\tau^2+C_1\right]^{-1/2}
\rmd \tau + C_2, \ \ m=-2.
\ear
 The variable $\tau$ changes on an interval, where
\bear{3.28}
S_m(\tau)= \eps
\left[
\frac{R_m(\tau)}{\tau R_m^{\prime}(\tau)}-1
\right]>0.
\ear
Finally, we get
\bear{3.29}
a=a_0 \tau^{\frac{4(1-h)}{3h(4-3h)}} S_m^{\frac{1}{3h}}(\tau)\\
\label{3.30}
\varphi=\frac{\sqrt{6}}{\kappa}{\rm sgn}(1-h)
\left\{
 \ln \left[\tau^{\frac{2(2-h)}{3h(4-3h)}}
 S_m^{\frac{1}{3h}}(\tau)\right]+ y_0 \right\},
 \ear
 where
 \bear{}
a_0= C^{\frac{4(1-h)}{3h(3h-4)}} (|V_0|/\rho_0)^{\frac{2(2-h)}{3h(3h-4)}},\
y_0=\ln
\left[
C^{\frac{2(2-h)}{3h(3h-4)}} (|V_0|/\rho_0)^{\frac{4(1-h)}{3h(3h-4)}}
\right].\nn
\ear
The variable $\tau$ and the cosmic time $t_c$ are related by
\bear{3.31}
\rmd  t_c^2= U_0
\tau^{\frac{4(2-3h)}{3h}}
S^{\frac{2(1-h)}{h}}_m(\tau)
\left[ R_m^{\prime}(\tau)\right]^2 \rmd \tau^2,
\ear
where
\bear{}
U_0= 4
\rho_0^{-\frac{4(1-h)^2}{h(3h-4)}}
|V_0|^{\frac{(2-h)^2}{h(3h-4)}}
C^{\frac{4(2-h)(1-h)}{h(3h-4)}}/(3h|3h-4|).\nn
\ear
So, this general solution is given in the parametrical form
by (\ref{3.29}),(\ref{3.30})
with $l$ and $m$ from (\ref{LM2}). The transition to the cosmic time
may be done by solving (\ref{3.31}).
\section{Properties of solutions}

 Here we study properties of the obtained exact solutions for positive
 potentials, though our solutions are valid for any sign of the potential.
  As the system is symmetrical under the time reflection
  $t\to -t$, without loss of generality we only consider expanding near  the
singularity cosmologies  with the Hubble parameter $H=\dot
x\exp(-\alpha)>0$.

We introduce the following notation for the relative energy densities
\bear{}
\Omega_{\rho}=\frac{\kappa^2\rho}{3H^2}, \
\Omega_{\varphi K}=\frac{\kappa^2(\rmd \varphi/\rmd t_c)^2}{6H^2},\
\Omega_{\varphi P}=\frac{\kappa^2V(\varphi)}{3H^2}.\nn
\ear
Due to the constraint (\ref{2.2}) the relative energy densities obey the
relation
\bear{}
\Omega_{\rho}+\Omega_{\varphi K}+\Omega_{\varphi P}=1.\nn
\ear
Also we introduce the scalar field barotropic parameter
\bear{}
\frac{p_{\varphi}}{\rho_{\varphi}}=\frac
{\frac{1}{2}\left(\frac{\rmd \varphi}{\rmd t_c}\right)^2    + V(\varphi)}
{\frac{1}{2}\left(\frac{\rmd \varphi}{\rmd t_c}\right)^2    -
V(\varphi)}.\nn
\ear

{\bf A1}. $0<\sigma<1$, $h=0$.\\
The general solution is given by Eqs.(\ref{3.2}),(\ref{3.3}) for
 $V_0(1-\sigma^2)>0,\ B\geq 0$. Near the initial singularity ($t_c\to +0$)
 the scale factor is in the main order $a\propto t^{1/3}_c$. There exists
the special solution for $A=-\sigma \sqrt{B}$ with $\Omega_{\rho}\to 1$ as
$t_c\to +0$. It corresponds to the fluid-dominated solution mentioned in
\cite{W98,W99,W02}.  It  gives
$p_{\varphi}/\rho_{\varphi}\to 1$ (stiff matter) as $t_c\to +0$.

For all remaining solutions we obtain
$\Omega_{\rho}\to 1 - \Omega_{\varphi K}^0,\ \
\Omega_{\varphi K}\to \Omega_{\varphi K}^0,\ \
\Omega_{\varphi P}\to 0$
as $t_c\to +0$,
where
$\Omega_{\varphi K}^0=(\sigma \sqrt{B}-A)^2/(\sqrt{B}-\sigma A)^2$.
Then the barotropic parameter $p_{\varphi}/\rho_{\varphi}$ tends to $-1$,
i.e the scalar field is vacuum-like (de Sitter) near the initial
singularity.

 There is an unique late-time attractor in the form of the scalar field
 dominated solution. It corresponds to $B=0$, and $\rho_0= 0$ in formulas
 (\ref{3.2}),(\ref{3.3}) presenting the general solution. The attractor
 may be written down for the cosmic time $t_c$

\bear{}
a=\tilde a_0 t_c^{1/(3\sigma^2)},\nn\\
\varphi=\frac{\sqrt{2/3}}{\kappa\sigma}
\left[\ln t_c + \ln\sqrt{\frac{3V_0\sigma^4}{1-\sigma^2}}\right].\nn
\ear
It is easy to see that for this solution
\bear{}
\Omega_{\rho}=0,\ \
\Omega_{\varphi K}=\sigma^2,\ \ \Omega_{\varphi P}=1-\sigma^2,\ \
\frac{p_{\varphi}}{\rho_{\varphi}}=2\sigma^2 - 1.\nn
\ear
 This attracting solution according to \cite{W02} may be called
 kinetic-potential scaling. For $\sigma^2<1/3$ all solutions provide
 the power law inflation at late times.

{\bf A2}. $1<\sigma$, $h=0$.\\
The general solution is given by (\ref{3.2}),(\ref{3.3}) for
$V_0(1-\sigma^2)<0,\ B> 0$. Behavior near the initial singularity is the
same
for both general and special solutions
with the only difference that the constant
$\Omega_{\varphi K}^0$ is the following
$\Omega_{\varphi K}^0=(\sigma \sqrt{B}+A)^2/(\sqrt{B}+\sigma A)^2$.

At the final stage of evolution as $t_c\to +\infty$ we obtain
$\Omega_{\rho}\to 1 - \Omega_{\varphi K}^f,\ \
\Omega_{\varphi K}\to \Omega_{\varphi K}^f,\ \
\Omega_{\varphi P}\to 0,\ \
p_{\varphi}/\rho_{\varphi}\to 1,\ \ a\propto t_c^{1/3}$,
where
$\Omega_{\varphi K}^f=(\sigma \sqrt{B}-A)^2/(\sqrt{B}-\sigma A)^2$.
Such behaviour of the scalar field, when it adopts the usual perfect
fluid equation of state and its energy-density scales with that of the
perfect fluid, is called scaling (or sometimes "tracking").

{\bf B} and {\bf C2} in regions {\bf Ia} and {\bf Ib}.\\
Behaviour near the initial singularity was described in
\cite{W98,W99,W02} using qualitative methods.  There
exists the fluid dominated solution with $a\propto
t_c^{2/[3(2-h)]}$ and $\Omega_{\rho}\to 1$ as $t_c\to +0$. All
remaining solutions describe the domination of the kinetic
contribution of the scalar field: $\Omega_{\rho}\to 0,\ \
\Omega_{\varphi K}\to 1,\ \ \Omega_{\varphi P}\to 0,\ \
p_{\varphi}/\rho_{\varphi}\to 1,\ \ a\propto t_c^{1/3}. $

The behaviour at late times is the same as one for {\bf A1}.

{\bf C1} and {\bf C2} in regions {\bf IIa} and {\bf IIb}.\\
Behaviour near the initial singularity is the same as in the previous case.
The late-time attractor for $h\in(0,h)$
is the special solution described by Eqs.(\ref{3.14})
and (\ref{3.15}). For these solution we have
\bear{}
\Omega_{\rho}=1 - \frac{2-h}{2\sigma^2},\
\Omega_{\varphi K}=\frac{(2-h)^2}{4\sigma^2},\
\Omega_{\varphi P}=\frac{h(2-h)}{4\sigma^2},\
\frac{p_{\varphi}}{\rho_{\varphi}}=1-h.\nn
\ear
This is a typical scaling behaviour.

One of the examples of this behaviour may be given explicitly:
the exact solution of the class {\bf C1} with $h=2$ and $\sigma=1/2$
for the positive potential
reads
with respect to the cosmic time
\bear{}
\fl a=a_0 \rme^{\sqrt{\Lambda/3}t_c}
\left\{ \frac{1-A\rme^{-\sqrt{3\Lambda}(t_c-t_c^0)}}
{1+A\rme^{-\sqrt{3\Lambda}(t_c-t_c^0)}}
\left[
\frac{ \sqrt{ 3\Lambda } } {2} (t_c-t_c^0) -
\frac{ 2A\left(1-A\rme^{ -\sqrt{ 3\Lambda }(t_c-t_c^0) }\right) }
{(1+A)\left(1+A\rme^{-\sqrt{3\Lambda}(t_c-t_c^0)}\right)}
\right]
\right\}
^{1/3},\nn
\ear
\bear{}
\fl
\varphi=
\frac{\sqrt{6}}{\kappa}
\ln
\left\{\frac{V_0}{\Lambda}
\left[
\frac{1+A\rme^{-\sqrt{3\Lambda}(t_c-t_c^0)}}
{1-A\rme^{-\sqrt{3\Lambda}(t_c-t_c^0)}}
\left(
\frac{ \sqrt{ 3\Lambda } } {2} (t_c-t_c^0) -
\frac{1-A}{1+A}
\right)-1
\right]
\right\}
^{1/3},\nn
\ear
where $\Lambda\equiv \rho_0$ is the cosmological constant. The solution
contains 3 integration constants: arbitrary $t_c^0$, positive $a_0$ and
A obeying $|A|<1$. The late-time attractor corresponds to $A=0$. For this
attracting solution we get:
$\Omega_{\rho}\to 1,\ \
\Omega_{\varphi K}\to 0,\ \
\Omega_{\varphi P}\to 0,\ \
p_{\varphi}/\rho_{\varphi}\to -1,\ \
a\propto t_c^{1/3}\exp{\sqrt{\Lambda/3}t_c},\ \
H\to \sqrt{\Lambda/3}$ as $t_c\to +\infty$.

\ack

This work was supported in part by the Russian Foundation for Basic
Research (Grant 01-02-17312) and DFG Project 436 RUS 113/678/0-1.
V.R.G and V.N.M. are grateful to colleagues of the university
of Konstanz for their hospitality.


\Bibliography{99}
\bibitem{Mel1}
Melnikov V N 1994
{\it Proc. Int. Conf. on Gravitation and Cosmology (Rio de Janeiro)}
(Singapore: Edition Frontieres) p147

\bibitem{Mel2} Melnikov V N  1995
{\it Multidimensional Gravitation and Cosmology II}
(Rio de Janeiro: CBPF-MO-002/95)

\bibitem{Mel3}
Melnikov V N 2002
{\it Exact Solutions in Multidimensional Gravity and Cosmology III}
(Rio de Janeiro: CBPF-MO-003/02)
\bibitem{FJ98}
Ferreira P G and Joice M 1998 \PR D {\bf 58} 023503\\
(Ferreira P G and Joice M 1997 {\it Preprint} astro-ph/9711102)

\bibitem{T01} Townsend P K 2001  Quintessence from M-theory
{\it Preprint}
hep-th/0110072

\bibitem{J95}  Chimento L P and  Jakubi A S 1995
Scalar field cosmologies with perfect fluid in Robertson-Walker metric
{\it Preprint}
gr-qc/9506015

\bibitem{W98}
Copeland E J, Liddle A R and  Wands D 1998 \PR D {\bf 57} 4686\\
(Copeland E J, Liddle A R and  Wands D 1997 {\it Preprint}
gr-qc/9711068)

\bibitem{B98}
Billyard A P, Coley A A and  van den Hoogen R J 1998
\PR D {\bf 58} 123501\\
(Billyard A P, Coley A A and  van den Hoogen R J 1998 {\it Preprint}
gr-qc/9805085)

\bibitem{W99}
van den Hoogen R J, Coley A A and Wands D 1999
\CQG {\bf 16} 1843\\
(van den Hoogen R J, Coley A A and Wands D 1999 {\it Preprint}
gr-qc/9901014)


\bibitem{W02} Heard I P C and Wands D 2002
Cosmology with positive and negative exponential potentials
{\it Preprint}
gr-qc/0206085


\bibitem{R01} Rubano C and Scudellaro P 2001
On Some exponential potentials for a cosmological scalar field as
quintessence
{\it Preprint}
gr-qc/01033335.

\bibitem{IM94}
Ivashchuk V D and  Melnikov V N 1994
{\it Int. J. Mod. Phys.} D {\bf 3}  795\\
(Ivashchuk V D and  Melnikov V N 1994 {\it Preprint} gr-qc/9403064)

\bibitem{GIM95}
Gavrilov V R, Ivashchuk V D and Melnikov V N 1995
{\it J. Math. Phys.} {\bf 36} 5829\\
(Gavrilov V R, Ivashchuk V D and Melnikov V N 1994
{\it Preprint} gr-qc/9407019)
\bibitem{GM98}
Gavrilov V R and Melnikov V N 1998
{\it Theor. Math. Phys.} {\bf 114} 454\\
(Gavrilov V R and Melnikov V N 1998
{\it Preprint} gr-qc/9801042)

\bibitem{PZ}
Polyanin A D and Zaitsev V F 1995
 {\it Handbook on Exact Solutions for Ordinary Differential Equations}
(Boca Raton: CRC Press)

\bibitem{ZP}
Zaitsev V F and  Polyanin A D 1994
{\it Discrete-Groups Methods for Integrating Equations of Nonlinear
Mechanics}
(Boca Raton: CRC Press)

\endbib
\end{document}